\documentclass[aps,prx,twocolumn,superscriptaddress]{revtex4-2}
\usepackage{amsmath}
\usepackage{amssymb}
\usepackage{color}
\usepackage{graphicx}
\usepackage{epstopdf}

\DeclareMathOperator{\arsinh}{arsinh}
\DeclareMathOperator{\artanh}{artanh}

\begin{document}
\title{Diffusioosmosis of electrolyte solutions in uniformly charged channels}

\author{Evgeny S. Asmolov}
\affiliation{Frumkin Institute of Physical Chemistry and Electrochemistry, Russian Academy of Sciences, 31-4 Leninsky Prospect, 119071 Moscow, Russia}
\author{Elena F. Silkina}
\affiliation{Frumkin Institute of Physical Chemistry and Electrochemistry, Russian Academy of Sciences, 31-4 Leninsky Prospect, 119071 Moscow, Russia}
\author{Olga I. Vinogradova}
\email[Corresponding author: ]{oivinograd@yahoo.com}
\affiliation{Frumkin Institute of Physical Chemistry and Electrochemistry, Russian Academy of Sciences, 31-4 Leninsky Prospect, 119071 Moscow, Russia}
\begin{abstract}
  When the concentration of electrolyte solution varies along the channel the forces arise that drag the fluid toward the higher or lower concentration region inducing a flow termed diffusio-osmotic. This article investigates a flow that emerges in channels with constant density of surface charge $\sigma$ and thin compared to their thickness electrostatic diffuse layers. An equation for the fluid flow rate $\mathcal{Q}$ is derived and used to describe analytically the flux of ions, and local potentials and concentrations. This equation, which allows to treat the diffusio-osmotic problems without tedious and time consuming computations, clarifies that the global flow rate is controlled only by the surface charge and concentration drop between the channel ends, and indicates that there exist generally two different values of $\sigma$ that correspond to a particular $\mathcal{Q}$. Our theory provides a simple explanation of the directions of the fluid flow rate and ionic flux depending on the surface charge and diffusivity of ions, predicts a non-linear concentration distribution along the channel caused by convection, and relates it to the local potential changes by a compact formula.  We also present and interpret the variations of the diffusio-osmotic velocity profiles and the apparent slip velocity along the channel and show that the latter is highly non-uniform and could even becomes alternating.  The relevance of our results for diffusio-osmotic experiments and for some electrochemistry and membrane science issues  is discussed briefly.
\end{abstract}

\maketitle

\section{Introduction}\label{sec:intro}

During recent decades, the pursuit of scale reduction has been extended to the fluidic domain and led to a rapid development of microfluidics and later nanofluidics~\cite{stone.ha:2004,squires.tm:2005,schoch2008,bocquet.l:2010}.
These have posed an issue of finding the more efficient mechanisms of generating flows in thin channels that exhibit huge hydrodynamic resistance
to a pressure-driven flow. One  avenue for driving flow on such small scales is to exploit hydrodynamic slip~\cite{vinogradova.oi:1999,bocquet.l:2007,vinogradova.oi:2011}. Another promising avenue is to employ  the so-called interfacially driven transport
phenomena that emerge in response to gradients of an electric potential, concentration of a solute, etc~\cite{anderson1989colloid}. The best known example is an electroosmosis, i.e. fluid flow induced by an applied (tangential) electric field.  Diffusioosmosis is an alternative, and less
explored, interfacially driven phenomenon that refers to flows under the gradient of a solute, for example a salt. An obvious advantage of diffusioosmosis is that it makes  possible to convert chemical (osmotic) energy into mechanical one, i.e. to a directed flow of solvent, and does not require an external energy supply.  The only source of energy is a concentration gradient of a solute, which is of much
interest for such applications as water purification technologies, energy generation, lab-on-a-chip devices,
and more~\cite{marbach2019,zhang2021}.

Despite its importance for several branches of surface physics and chemistry, as well as for applications, diffusioosmosis caused by a salinity gradient has so far received much less attention than electroosmosis. The successful understanding of its origin, due to \citet{deryagin1961} and \citet{prieve1984motion}, was an important achievement of 20th century colloidal hydrodynamics. Anticipating the discussion to follow in later, these authors have been the first to realize that in the case of electrolyte solutions there exist two contributions that originate the steady-state diffusioosmosis. The first one, termed chemiosmotic, is associated with the osmotic pressure gradient in the electrostatic diffuse layers (EDLs) formed in the neighborhood of the charged walls that extend to distances of the order of Debye length $\lambda_D$ of the bulk electrolyte solution. The second contribution is associated with the electroosmosis under a spontaneously arising electric field that ensures the same propulsion velocity of ionic species of  different diffusivity (which is equivalent to saying that the electric current vanishes). While the chemiosmotic flow is always toward the lower electrolyte concentration, the electroosmosis could be directed to any side depending on the difference in diffusivity of anions  and  cations, as well as on the sign of the surface charge.

During the past decades this scenario of  diffusioosmosis  has become widely accepted for electrolyte solutions. Nevertheless,  the procedure of calculating the velocity of diffusio-osmotic flow is still beset with difficulties, especially in the case of channels of a finite thickness. To what extent can notions of a flow near a single wall be employed
in confined systems? For example, should  the condition of zero current be applied outside of the EDLs or differently? How do the concentration and emerging electric field vary along the channel? What is the velocity profile of a confined
fluid and how does this depend upon  the nature of the
confining walls and concentration drop? Since the answers to these questions still cannot be obtained from experiment, it is not surprising that fundamental understanding of diffusioosmosis in micro- and nanochannels did not begin to emerge until quite recently.
There is now a growing literature describing attempts  to provide a satisfactory quantitative theory of diffusioosmosis in a slightly nonuniform salt solution. We mention below what we believe are the more relevant contribution.

A large fraction of theoretical papers deal with an idealized case of a single, infinitely long, wall. \citet{deryagin1961} appear to have
been the first to address the issue of diffusioosmotic flow caused by a gradient in the concentration of a simple salt and to argue that
a spontaneous electric field emerges (to cancel the electric current far from the walls, i.e. beyond the EDLs, out), which, in turn, gives rise to a supplementary (electroosmotic) contribution to the total flow. These authors  derived an equation that relates this emerging field with the ion diffusion coefficients, the local concentration of salt and its gradient in the bulk, but do not present any detailed results for a diffusio-osmotic velocity.
A more systematic treatment of diffusioosmosis was contained in remarkable papers published by~\citet{prieve1984motion} and
\citet{anderson1989colloid}. The calculations [based on a system of the Nernst-Planck equations for the concentration of ion species, Poisson equation for the electric potential generated by inhomogeneous charge distribution and Stokes equation for fluid flow] led to an analytical solution for the local velocity of diffusio-osmotic plug flow far from the wall, i.e. outside of the diffuse layer. This outer velocity $U_s$ was termed the diffusioosmotic slip velocity since macroscopically it appears
that the liquid slips over the surface.  The authors~\cite{prieve1984motion} have shown that the local parameters determining both contributions to this (apparent) slip velocity are the  surface potential, bulk concentration of salt and its gradient, but have not tried to derive them self-consistently.
This was taken up later  by \citet{keh2005} who obtained a solution for $U_s$ by postulating a linear variation  in bulk concentration along the wall. However, no attempts have been made to  explain how to impose such a constant gradient externally and/or whether could it naturally occur.

\begin{figure}[h]
\includegraphics[width=0.9\columnwidth]{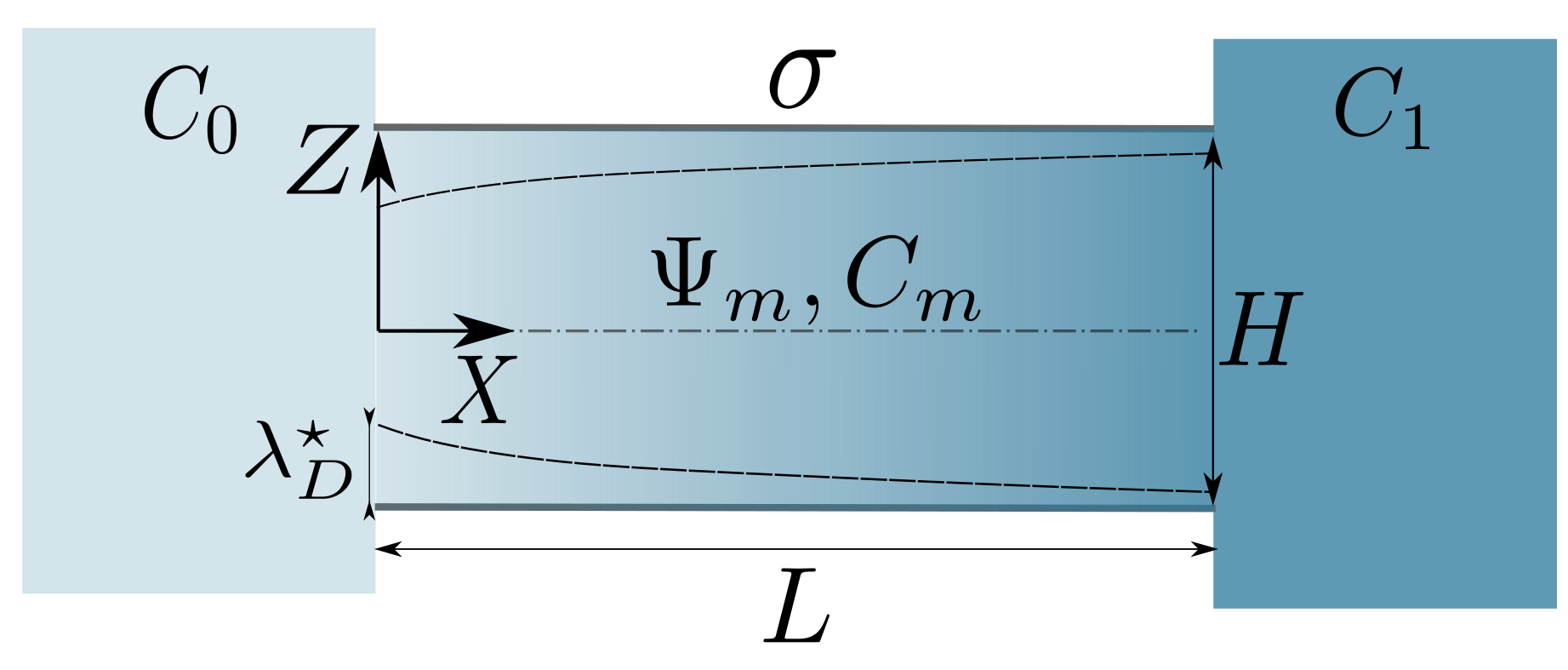}
\caption{Sketch of the microchannel of thickness $H$, length $L
\gg H$ and constant surface charge density $\sigma$ that connects the ``fresh'' (left) and ``salty'' (right) bulk electrolyte reservoirs of
concentrations $C_{0}$  and $C_{1}$. The extension of electrostatic
diffuse layers, which is of the order of the Debye length $\protect\lambda_D \ll H$ and takes its upper value $\lambda_D = \lambda_D^{\star}$ at $X=0$ by reducing along the channel. The electroneutral central area is of ``bulk'' concentration $C_m$ and potential $\Psi_m$, both depend on $X$.  }
\label{fig:sketch}
\end{figure}

During the last years several theoretical papers have been concerned with the diffusio-osmotic flow in a planar channel of finite thickness $H$ and length $L$ connecting two reservoirs of different salinity (as sketched in Fig.~\ref{fig:sketch}).
\citet{ma.hc:2006} and \citet{keh.hj:2016} address themselves to the problem of the diffusio-osmotic flow in a thick  ($H \gg \lambda_D$) channel. These authors  performed detailed calculations of $U_s$ assuming that the electric current has to vanish locally, i.e. at any point of the channel. Later, however, \citet{jing2018} argued that in the case of channel of finite thickness it would be more appropriate to apply the zero net (i.e. averaged over a cross-section) current condition. Note that neither paper attempted to relax the simplifying assumption about a
constant concentration gradient along the channel or
address itself to the issue of its calculation.
While this conventional assumption
appears to be reasonable for a small difference between bulk concentrations of reservoirs, it is by no means obvious in the most interesting and practically relevant case of a finite concentration drop. In  any  event,  it  seems  clear  that the variation in the concentration gradient along the channel and local concentrations are established self-consistently and have to be calculated. \citet{ault.jt:2019} performed the calculations for a constant surface potential channel  and found that the local concentration inside varies non-linearly. These authors, however, made an assumption that the fluid flow rate is initially prescribed and have not attempted to derive it. We are unaware of any theoretical work that addressed this issue.

Numerical calculations might be expected to shed some light on diffusio-osmotic phenomena, and indeed in prior investigations numerical approaches have mostly been followed. The diffusio-osmotic velocity has been computed from a solution to a system of partial differential equations   establishing linear relationships between local fluxes averaged
across channel cross-sections (the current, the flow rate of fluid and fluxes of ions) and
driving forces (electric field, gradients of salt concentration and
pressure)~\cite{fair1971,sasidhar1981,peters2016,jing2018}. The significance of~\citet{prieve1984motion} paper does not seem to be recognized in these publications.
Most subsequent attempts at improvements on numerical calculations of the diffusio-osmotic flow~\cite{qian2007,ryzhkov2016,chanda2022} have also either ignored or failed to make direct connection with earlier analytical results~\cite{prieve1984motion,anderson1989colloid,ma.hc:2006}.

Given the current upsurge of the interest in the diffusio-osmotic phenomena and their applications it would seem appropriate to bring more theoretical analysis to bear on this problem. In this paper we present some results of a study of the steady-state diffusio-osmotic flow in the channel of a constant surface charge density assuming thin diffuse layers compared to its thickness, but do not make any simplifying assumptions about the magnitudes of the surface potential and the  concentration drop between the reservoirs. Our theory is based on an approach introduced originally by~\citet{prieve1984motion} in their study of a local diffusio-osmotic slip near a single wall. We shall see that the extension of the theory to the
two-wall case provides new insight into the physics of diffusioosmosis  yielding useful (approximate) analytical results, as well as suited to numerical work. These features are especially advantageous when one is attempting to calculate the diffusio-osmotic velocity at a large concentration drop.

Our paper is arranged as follows: In Sec.~\ref{sec:model} we define our system and formulate the governing equations. Section~\ref{sec:potential} discusses the perturbation of the electrostatic potential due to diffuse layers. Derivations of equations for the local surface potentials are given. In  Sec.~\ref{sec:velocity} the extension of the~\citet{prieve1984motion} approach to two walls is described and the equation for the local diffusio-osmotic mobility profile  is derived. The differential equation for the local diffusio-osmotic slip is also formulated here. Section~\ref{sec:fluxes} describes the theory of a global diffusio-osmotic flow in the channel. In Sec.~\ref{sec:general} some general considerations concerning the flow rate of fluids and ionic flux in the channel are given. Here we also argue  that the zero current condition should be imposed to its average (over the cross-section) value.
The procedure for calculating the concentration and potential in the electroneutral (``bulk'') part of the channel is described in Sec.~\ref{sec:equations}. After establishing a relationship between the distributions of potential and concentration  in the ``bulk'' part of the channel, we obtain an expression that relates the latter to the flow rate of fluid, which can be easily found by integrating a simple function  we derive. Since the calculated concentration distributions along the channel are, in general, nonlinear, to provide the same flow rate in every cross-section the pressure gradient arises naturally in our analysis by giving rise to a supplementary flow.  In Sec.~\ref{sec:slip} the form of the velocity profiles is described; this depends on the locus of the given cross-section since the pressure gradient is alternating. We also consider the apparent slip velocity on the walls and discuss its variations along the channel in different situations. Finally, we propose simple and compact analytical approximations for the diffusio-osmotic slip velocity, which apply when the surface potential becomes small. We conclude in Sec.~\ref{sec:concl} with a discussion of our results and their relevance.

\section{Model and governing equations}\label{sec:model}

We consider a system sketched in Fig.~\ref{fig:sketch}. A planar channel of thickness $H$ and length $L \gg H$ is in contact with low (left) and  high (right) salinity  reservoirs of symmetric 1:1 salt solutions of concentrations (number densities of salt molecules) $C_0$ and $C_1$ at standard temperature $T$. Both solutions are of the same dynamic viscosity $\eta $ and permittivity $\varepsilon$.

It is convenient to place the origin of coordinates in the midplane of the channel at a junction with the low-salinity bath. The $Z$-axis is aligned
across the slit, so the two walls  are located at $Z=\pm H/2$, and for a symmetric channel it is enough to consider $0\leq
Z\leq H/2$. The $X$-axis is defined along the channel, and $0\leq X\leq L$.

For a given thickness $H$ and length $L$ the  solution inside the slit will adopt the configuration (i.e. distributions of the number density of the ionic species $C^{\pm}$, electrostatic potential $\Psi$ and hydrostatic pressure $P$) that minimizes the energy dissipation, i.e. corresponds to steady-state diffusio-osmotic flow with the velocity $\mathbf{U}$.

The scenario of a stationary diffusio-osmosis can be understood as follows. When we connect low and high salinity reservoirs,
cations and anions begin to diffuse along the channel towards the ``fresh'' bath, but their diffusion coefficients $D^{+}$ and $D^{-}$ are generally unequal.
Ions with larger diffusion coefficients migrate faster, but a local ionic concentration can approach the steady-state if and only if their fluxes
at any cross-section are equal. This is equivalent to saying that no electric current occurs throughout the channel. To provide this, a tangential
electric field should emerge to accelerate slower ions and retard faster ones. This field gives rise to
an electroosmotic flow along the channel. Simultaneously, the osmotic pressure gradient occurring within the EDLs and balanced by the viscous
stress induces a chemiosmotic flow from the ``salty'' to ``fresh'' reservoir. In addition, a hydrostatic pressure distribution along the channel
emerges. These contributions are superimposed, and the stationary state of resulting diffusio-osmotic flow is described by the system of several
governing equations specified below.

The Nernst-Planck (or convection-diffusion) equation describes the conservation  of ionic
species at each point $(X,Z)$ inside the channel
\begin{equation}
\mathbf{\nabla }\cdot \mathbf{J}^{\pm }=0,  \label{NPH}
\end{equation}
where the ionic fluxes $\mathbf{J}^{\pm }$ of cations and anions are given by
\begin{equation}
\mathbf{J}^{\pm }=C^{\pm }\mathbf{U}+D^{\pm }\left( -\mathbf{\nabla }C^{\pm
}\mp \dfrac{e}{k_{B}T}C^{\pm }\mathbf{\nabla }\Psi \right).
\label{jd}
\end{equation}%
Here $e$ is the elementary
positive charge and $k_{B}$ is the Boltzmann constant. The first term in Eq.~\eqref{jd} is associated with the convective flux of ions induced by the flow, the second refers to  the diffusive drift relative to a solvent, and the third one is due to migration of ions in the emerging electric field.

The relation between the potential $\Psi$ and the charge density $\rho$  is
given by the Poisson equation:
\begin{equation}
\Delta \Psi =-\frac{\rho }{4\pi \varepsilon }=-\frac{e\left(
C^{+}-C^{-}\right) }{4\pi \varepsilon }.  \label{PEq}
\end{equation}%

The fluid flow satisfies the Stokes equations,
\begin{eqnarray}
\mathbf{\nabla \cdot U} &=&0\mathbf{,\quad }  \label{cont} \\
\eta \Delta \mathbf{U}-\mathbf{\nabla }P &=&\rho \mathbf{\nabla }\Psi .
\label{mom}
\end{eqnarray}%

It is clear that the theory for the diffusio-osmosis requires, as input, the hydrodynamic and electrostatic boundary conditions at the channel
walls. Here we consider no-slip ($\mathbf{U = 0}$ at $Z = \pm H/2$) nonconducting surfaces of charge density $\sigma$, which is constant, i.e. independent
on $X$. Rather than using $\sigma$ explicitly we here describe the surfaces by the Gouy-Chapman length
\begin{equation}
\ell _{GC}=\dfrac{e}{2\pi \sigma \ell _{B}},  \label{eq:LGC}
\end{equation}%
where $\ell _{B}=\dfrac{%
e^{2}}{\varepsilon k_{B}T}$ is the Bjerrum length. The Gouy-Chapman length is inversely proportional to the surface charge density (and may be positive or negative depending on its sign).

The partial differential equations \eqref{NPH}-\eqref{mom} we have described above are generic and apply at any point $(X,Z)$ of the channel of an arbitrary thickness. They can generally be solved only numerically. However, below we show that approximate analytical results can be obtained in the  limit of thick channel, which, essentially, implies that a compensating charge of the opposite sign and equal magnitude staying in the neighborhood of the charged walls is confined in very thin electrostatic diffuse layers (EDLs) near the walls, while an extended central region of the channel  is then, to the leading order, electro-neutral (``bulk'') in any cross-section. This implies that at any $X$ it is admissible to divide the thick channel into a surface (inner) and a bulk  (or outer) regions. The  concentration $C_m$ and potential $\Psi _{m}$ in the bulk region are then independent on $Z$ and vary slowly in the $X-$ direction.

The local EDL thickness is of the order of the Debye screening length at a given cross-section and can be defined as
\begin{equation}\label{eq:DL1}
 \lambda _{D} = \left[8\pi \ell _{B}C_m\right] ^{-1/2} \propto C_m^{-1/2},
\end{equation}
Clearly, $\lambda _{D}$ reduces on increasing  $X$ since the concentration augments. The upper value of $\lambda_D$ is thus attained at $X=0$. In this cross-section $C_m = C_0$ that yields
\begin{equation}\label{eq:DL2}
 \lambda^{\star}_{D} = \left[8\pi \ell _{B}C_0\right] ^{-1/2}.
\end{equation}
Therefore, to fulfil the thick channel condition, it is enough to require $\lambda_{D}^{\star} \ll H$.

Note that by analyzing the experimental data it is more convenient to use the concentration $\mathcal{C}[\rm{mol/l}]$, which is related to $C$ as $C \simeq N_A \times 10^3 \times \mathcal{C}$, where $N_A$ is Avogadro's number. The Bjerrum
length of water at $T \simeq 298$~K is equal to about $0.7$ nm leading to
\begin{equation}\label{eq:DLength}
  \lambda^{\star}_D [\rm{nm}] \simeq \frac{0.305 [\rm{nm}]}{\sqrt{\mathcal{C}_0 [\rm{mol/l}]}}.
\end{equation}
Upon increasing $\mathcal{C}_0$ from $10^{-6}$ to $1$ mol/l the screening length reduces from about 300 down to 0.3 nm. This implies that if, say,
we take $H=100$ nm, then the thick channel limit is expected when $\lambda^{\star}_D \leq 10$ nm, i.e. provided that in the ``fresh'' bath $\mathcal
{C}_0 \geq 10^{-3}$ mol/l. In all calculations below we fix this value of $H$ and use $\mathcal{C}_0 = 10^{-3}$ mol/l.

In the thick channel limit the local electrostatic potential in the slit is given by
\begin{equation}\label{eq:PSI}
 \Psi (X,Z) = \Psi_m (X) + \Phi (X,Z).
\end{equation}
Here the ``bulk'' term ($\Psi_m$) is supplemented by a ``surface'' term ($\Phi$) that represents the perturbation due to diffuse layers. Eq.~\eqref{eq:PSI} implies that the emerging electric field that produce electroosmotic flow is additively superimposed upon the field of the EDL. Later we shall see that for our configuration $\Phi$ is independent on the fluid flow.

To construct the solution of the system of Eqs.~\eqref{NPH}-\eqref{mom} it is convenient to define the dimensionless coordinates
\begin{equation*}
 x = \frac{X}{L},\quad z=\frac{2Z}{H}
\end{equation*}
that vary from $0$ to $1$. The dimensionless potentials are defined as
\begin{equation*}
\psi = \dfrac{e\Psi }{k_{B}T},\quad \phi = \dfrac{e\Phi }{k_{B}T}.
\end{equation*}
We also introduce the dimensionless variables~\cite%
{saville1977}
\begin{eqnarray*}
	\mathbf{j}^{\pm } &=&\mathbf{J}^{\pm }\frac{2L}{C_{0}\left(
		D^{+}+D^{-}\right) },\quad c^{\pm }=\frac{C^{\pm }}{C_{0}}, \\
	\mathbf{u} &\mathbf{=}&\mathbf{U}\dfrac{4\pi \eta \ell_B L}{k_{B} T},\quad p=P\dfrac{\pi \ell_B H^{2}}{k_{B} T}
\end{eqnarray*}%

\section{Electrostatic potential}\label{sec:potential}

Using the dimensionless variables the Nernst-Planck \eqref{NPH} and the continuity \eqref{cont} equations can be rewritten as
\begin{equation}
\frac{H}{2L}\partial _{x}j_{x}^{\pm }+\partial _{z}j_{z}^{\pm }=0,
\label{npz}
\end{equation}%
\begin{equation}
\frac{H}{2L}\partial _{x}u_{x}+\partial _{z}u_{z}=0.  \label{cont_z}
\end{equation}%
For a long channel, $H/L \ll 1,$ the leading-order terms in Eqs.~\eqref{npz} and \eqref{cont_z} involve only derivatives with respect to $z$.

Applying the impermeability condition
\begin{equation}
j_{z}^{\pm }\left( x,\pm 1\right) =u_{z}\left( x,\pm 1\right) =0,
\label{imper}
\end{equation}%
we conclude that the normal fluxes $j_{z}^{\pm }$ and velocity
$u_{z}$ are of the order of $H/L$, i.e. small. Since
$j_{z}^{\pm }\simeq 0$, the ion
concentrations obey local Boltzmann distributions at any cross-section \cite%
{fair1971,peters2016}:
\begin{equation}
c^{\pm }=c_m\left( x\right) \exp \left( \mp \phi \right) ,  \label{cpm}
\end{equation}%
where
\begin{equation}
\phi =\psi (x,z)-\psi_m\left( x\right),   \label{fd}
\end{equation}%
which is Eq.~\eqref{eq:PSI} rewritten in dimensionless form.
Here the midplane (``bulk'') concentration $c_m$ and the potential $\psi_m$
vary only in $x$ direction. The boundary conditions for $c_m$ are
\begin{equation}
c_m\left( 0\right)  =1,\quad c_m \left( 1\right) =c_{1}=C_{1}/C_{0}.  \label{c0}
\end{equation}

We also set
\begin{equation}
\psi_m\left( 0\right)  =0, \label{psi0}
\end{equation}%
but the value of $\psi_m\left( 1\right) $ that ensures the fulfillment of
the condition of zero current is initially unknown and has to be determined.

The potential $\phi $ satisfies the Poisson-Boltzmann equation at each
cross-section,
\begin{equation}
\partial _{zz}\psi =\partial _{zz}\phi =c_m\left( x\right) \lambda ^{-2}\sinh
\phi ,  \label{PB}
\end{equation}%
\[
\lambda = \frac{2 \lambda_{D}^{\star}}{H}.
\]%

To integrate Eq.~\eqref{PB} we impose two electrostatic boundary conditions. Symmetry of the
channel dictates that
\[
\partial _{z}\psi \left( x,0\right) =\partial _{z}\phi \left( x,0\right) =0.
\]%
Another condition requires a constant surface charge density of the walls, which is equivalent to a constant gradient of the surface potential.
\begin{equation}
\partial_{z}\psi \left( x,1\right) =\partial _{z}\phi \left( x,1\right) =%
\frac{H}{\ell _{GC}}.  \label{cc}
\end{equation}%
However, the surface potential itself varies with $x$ since a local Debye length scales with $c_m^{-1/2}.$

The first integration of the Poisson-Boltzmann equation \eqref{PB} from $0$ to $z$ leads to
\begin{equation}
\frac{\left( \partial _{z}\phi \right) ^{2}}{2}=c_m \lambda ^{-2}\left( \cosh
\phi -1\right) =2c_m \lambda ^{-2}\sinh^{2}\left(\frac{ \phi}{ 2}\right)   \label{op}
\end{equation}%
or
\begin{equation}
 \partial _{z}\phi =  2 \lambda ^{-1} c_m^{1/2}\sinh\left(\frac{ \phi}{ 2}\right)   \label{op2}
\end{equation}%
This equation is identical to that for a single wall, although the boundary conditions were different [in the single wall problem both $\phi$ and $\partial _{z}\phi$ vanish far away from the wall].

From Eqs.~\eqref{cc} and \eqref{op2} it follows that
the relation between the surface potential and charge is given
by
\begin{equation}
\phi _{s} =  2\arsinh\left( \frac{\lambda _{D}^{\star}}{c_m^{1/2}\ell
_{GC}}\right),  \label{eq:pot-charge_hs}
\end{equation}
which can be regarded as a direct analog of the Grahame equation for a single wall~\cite{israelachvili.jn:2011}.

Note that $\lambda _{D}^{\star}/(c_m^{1/2}\ell_{GC})$ plays a role of an effective surface charge density that reduces on increasing $x$.
When it is small, Eq. (\ref{eq:pot-charge_hs}) can be
written as%
\begin{equation}
\phi_{s}\left( x\right) \simeq \frac{2\lambda^{\star}_{D}}{c_m^{1/2}\ell _{GC}} = \frac{\lambda H}{c_m^{1/2}\ell _{GC}}\ll 1.
\label{fs_DB}
\end{equation}%

If the effective surface charge is large, the surface potential depends on it weakly logarithmically:
\begin{equation}\label{fs_DB2}
\phi_{s}\left( x\right) \simeq \pm 2 \ln \left[\dfrac{2 \lambda _{D}^{\star}}{c_m^{1/2}|\ell_{GC}|}\right] = \pm 2 \ln \left[\dfrac{ \lambda H}{c_m^{1/2}|\ell_{GC}|}\right]
\end{equation}
The choice of sign in \eqref{fs_DB2} depends on whether the walls are positively or negatively charged. The plus sign must be taken for a positive $\ell_{GC}$, and vice versa.

Performing the integration in \eqref{op2} yields an exact analytical solution for the disturbance potential
 \begin{equation}\label{eq:PBSWexact}
 \phi (x,z) = 4 \artanh \left[e^{(z-1)c_m^{1/2}/\lambda}\tanh \left(\dfrac{\phi_s}{4}\right)\right],
\end{equation}
where $\phi_s$ is given by \eqref{eq:pot-charge_hs}.

\begin{figure}[h]
\begin{center}
\includegraphics[width=1\columnwidth]{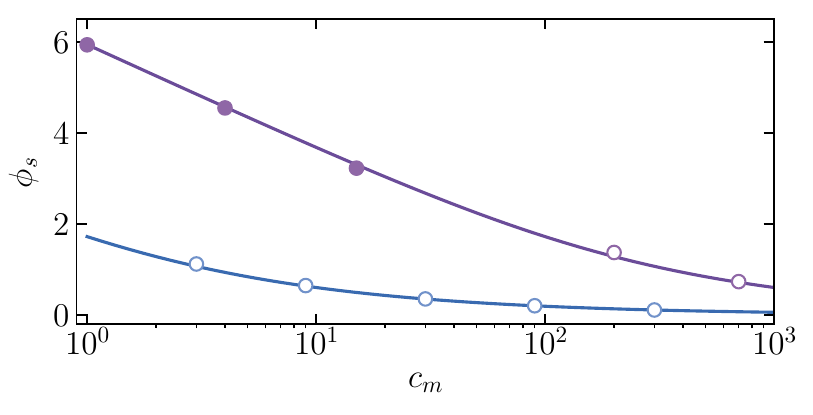}
\end{center}
\caption{Surface potential $\phi_s$ as a function of $c_m$ computed
using $\ell_{GC} = 1$ and 10 nm (solid curves from top to bottom) for $\mathcal{C}_{0} = 10^{-3}$
mol/l and $\mathcal{C}_{1} = 1$ mol/l. Open and filled circles show calculations from Eqs.~\eqref{fs_DB}  and \eqref{fs_DB2}. }
\label{fig:pot-charge}
\end{figure}

A plot of $\phi_s (c_m)$ is given in Fig.~\ref{fig:pot-charge}. For this numerical example we use $H = 100$ nm and perform integration of Eq.~\eqref{PB} [using an approach described before~\cite{silkina.ef:2019,herrero.c:2024}]. The calculations from~\eqref{eq:pot-charge_hs} are also made, but not shown since they fully coincide with numerical results. It can be seen that on increasing $c_m$ (and hence $x$) the surface potential reduces and that at a given ``bulk'' concentration it is always higher for surfaces of smaller $\ell_{GC}$.
Also included in Fig.~\ref{fig:pot-charge} are approximate surface potentials predicted by  Eqs.~\eqref{fs_DB}  and \eqref{fs_DB2}. The fits are quite good for small and large $\phi_s$, correspondingly, which is expected since $\lambda_D^{\star} \simeq 10$ nm is much smaller than $H$.

\section{Local diffusio-osmotic slip}
\label{sec:velocity}

From (\ref{PB}) it follows that $x-$momentum equation (\ref{mom}) can be transformed into a linear differential equation~\cite{prieve1984motion,jing2018}
\begin{equation}
\partial _{zz}u_{x}=\partial _{x}p-\frac{c_{m}\sinh \phi }{\lambda ^{2}}%
\partial _{x}\psi _{m}+\frac{\cosh \phi -1}{\lambda ^{2}}\partial _{x}c_{m}.
\label{dzzu}
\end{equation}%
Here the hydrodynamic (pressure-driven) term is supplemented by an
electroosmotic one and a chemiosmotic contribution.

Integrating this equation twice with respect to $z$ and applying the no-slip $%
u_{x}\left( 1\right) =0$ and symmetry $\partial _{z}u_{x}\left( 0\right) =0$
boundary conditions we derive
\begin{equation}
u_{x}=m_{h}\partial _{x}p+m_{e}\partial _{x}\psi _{m}+m_{c}\partial
_{x}c_{m},  \label{vel}
\end{equation}%
where the local hydrodynamic and electroosmotic mobilities are given by
\begin{equation}
m_{h}=-\frac{1-z^{2}}{2}, \label{uh}
\end{equation}
\begin{equation}
m_{e}=-\phi +\phi _{s}.  \label{ueo}
\end{equation}
The latter equation implies that the sign of $m_e$ is defined solely by that of $\phi_s$. An electroosmotic mobility will be positive for positively charged surfaces and negative for negatively charged.

\begin{figure}[h]
\begin{center}
\includegraphics[width=1\columnwidth]{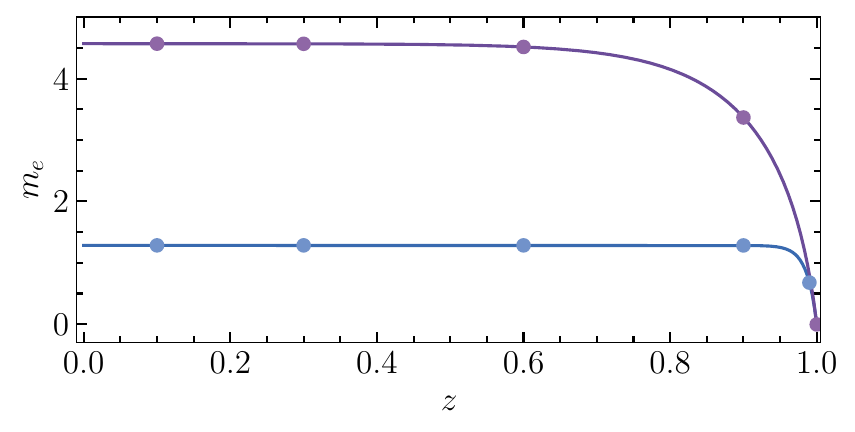}
\end{center}
\caption{Electroosmotic mobility profiles computed for cross-sections of $\mathcal{C}_m = 4 \times 10^{-3}$ and $0.2$ mol/l (solid curves from top to bottom) for the upper curve in Fig.~\ref{fig:pot-charge}. Circles show $m_e$ calculated from Eq.~\eqref{ueo} using \eqref{eq:pot-charge_hs} and \eqref{eq:PBSWexact}. }
\label{fig:pot-profile}
\end{figure}

Figure~\ref{fig:pot-profile} shows the electroosmotic mobility profiles with the same parameters as for the upper curve in
Fig.~\ref{fig:pot-charge}. The calculations are made from Eq.~\eqref{ueo} [by substituting $\phi_s$ and $\phi$ given by equations \eqref{eq:pot-charge_hs} and  \eqref{eq:PBSWexact}] for cross-sections of $\mathcal{C}_m = 4 \times 10^{-3}$ and $0.2$ mol/l, which corresponds to $c_m = 4$
and $200$. It can be seen that the (positive) electroosmotic mobility in the central region shows a distinct plateau and varies only
inside the EDLs to vanish at the walls. It is important to note that the existence of the plateau regions indicates that $\partial _{zz}\phi
 = 0$, i.e. the left-hand side of \eqref{PB} vanishes. The right-hand side is the charge, and hence this vanishes too. Thus, the central part of 
the channel is indeed electro-neutral. On reducing $\mathcal{C}_m$ the plateau height
decreases, which reflects the variation in $\phi_s$, but
remains finite for both specimen examples presented in Fig.\ref{fig:pot-profile}.

The local chemiosmotic mobility may, in turn, be determined from
\begin{equation}
m_{c}=-\frac{1}{\lambda ^{2}}\int_{z}^{1}\int_{0}^{y}\left[ \cosh \phi
\left( s\right) -1\right] ds dy  \label{udo}
\end{equation}%
using Eq.~\eqref{op}. Performing the first integration yields
\begin{eqnarray*}
\int_{0}^{y}\left( \cosh \phi -1\right) ds &=&\lambda
c_{m}^{-1/2}\int_{0}^{y}\sinh \left( \phi /2\right) d\phi  \\
&=&2\lambda c_{m}^{-1/2}\left[ \cosh \left( \phi /2\right) -1\right] .
\end{eqnarray*}%
The second integral can then be calculated as follows
\begin{eqnarray}
m_{c} &=&-c_{m}^{-1}\int_{\phi }^{\phi _{s}}\frac{\cosh \left( \phi
/2\right) -1}{\sinh \left( \phi /2\right) }d\phi  \\
&=&-4c_{m}^{-1}\left\{ \ln \left[ \cosh \left( \phi _{s}/4\right) \right]
-\ln \left[ \cosh \left( \phi /4\right) \right] \right\} .  \label{udo2}
\end{eqnarray}%
Since the expression in curly brackets is positive for any $\phi_s \neq 0$, we conclude that the chemiosmotic mobility can only be negative.

In the case of low surface potentials, $\left\vert \phi _{s}\right\vert \leq
1$, one can use the approximation $\ln \left[ \cosh \left( \phi /4\right) %
\right] \simeq \phi ^{2}/32$. Equation~\eqref{udo2} then reduces to
\begin{equation}
m_{c}\simeq -\frac{\phi _{s}^{2}-\phi ^{2}}{8c_{m}}\ll 1  \label{mdo_sm}
\end{equation}%
Thus, the chemiosmotic mobility turns out to be small compared to the
electroosmotic one [given by Eq.~\eqref{ueo}].

\begin{figure}[h]
\begin{center}
\includegraphics[width=1\columnwidth]{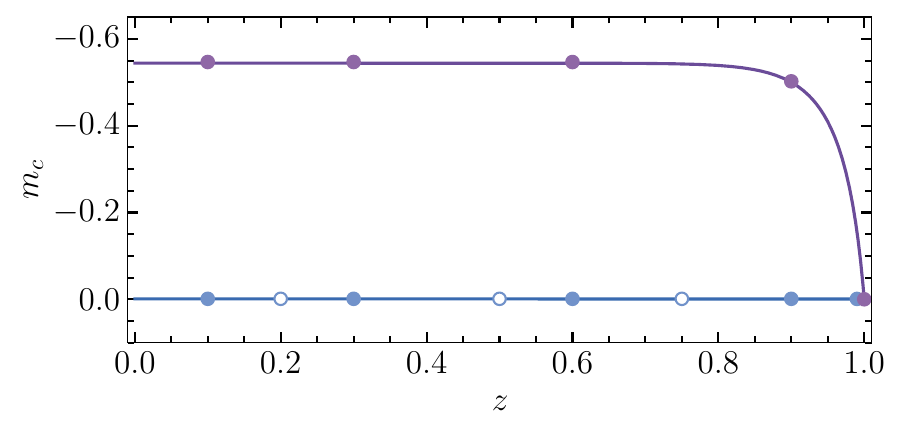}
\end{center}
\caption{Chemiosmotic mobility $m_{c}$ as a function of $z$ computed with
the same parameters as in Fig.~\ref{fig:pot-profile}. Filled and open circles are calculations from Eqs.~\eqref{udo2} and \eqref{mdo_sm}. }
\label{fig:mdo}
\end{figure}

Figure~\ref{fig:mdo} includes curves for the local chemiosmotic mobility obtained by numerical integration of Eq.~\eqref{udo} using $\phi$ determined by Eq.~\eqref{eq:PBSWexact}. The calculations are made for the same parameters as in Fig.~\ref{fig:pot-profile}. The shape of these curves for (negative) $m_c$ is similar to that for electroosmotic mobility, so that we do not discuss it in detail, but note
that for both cross-sections $|m_c|$ is smaller than unity (and significantly below $m_e$). Both curves now are well fitted by Eq.~\eqref{udo2}.
Also included in Fig.~\ref{fig:mdo} is the chemiosmotic mobility calculated from asymptotic Eq.~\eqref{mdo_sm}. The fit is good for the lower curve
($\mathcal{C}_m = 0.2$ mol/l), but clearly, the exact results for the upper curve ($\mathcal{C}_m = 4 \times 10^{-3}$ mol/l) should be irreconcilable
with Eq.~\eqref{mdo_sm} since $\phi_s \simeq 4.5$ is finite. Indeed, there is some discernible discrepancy in the direction of greater $|m_c|$ (not
shown).

The above equations for mobilities apply at any point of the channel, but since
$\phi $ vanishes outside the EDLs, we might argue that in
the central electroneutral region Eqs.~\eqref{ueo} and \eqref{udo2} reduce to
\begin{equation}
m_{e}\simeq \phi _{s},\ m_{c}\simeq -\frac{4\ln \left[ \cosh \left( \phi
_{s}/4\right) \right] }{c_{m}}.  \label{eq:mthick}
\end{equation}

The sum of
electro- and chemiosmotic mobilities, Eqs. (\ref{ueo}) and (\ref{udo2}),
corresponds to diffusio-osmotic plug flow with dimensionless slip velocity,
\begin{equation}
u_{s}=\phi _{s}\partial _{x}\psi
_{m}-4\frac{\partial _{x}c_{m}}{c_{m}}\ln \left[ \cosh \left( \frac{\phi _{s}}{4}\right) \right] ,  \label{ui}
\end{equation}%
where given by (\ref{eq:pot-charge_hs}) surface potential $\phi _{s}$  is salt-dependent.  It becomes evident that, if $\phi _{s}=0$, both
terms vanishes, so the diffusio-osmotic flow does not emerge. We remark that
it also cannot be generated if electro- and chemiosmotic contribution cancel
each other. It is of considerable interest to determine whether and when
this could happen. In all other situations $u_s \neq 0$, and at this stage it is not clear will it be positive or negative for some special configuration. While the chemiosmotic contribution can only be negative (since the derivative $\partial _{x}c_{m}$ is positive), the electro-osmotic term can be of any sign depending on $\partial _{x}\psi
_{m}$, which is still unknown. One further comment should be made. Eq.~(\ref{ui}) is non-linear in $c_{m}$. This implies that in the general case the diffusio-osmotic slip velocity varies along the channel.

We recall that the above treatment corresponds to cross-sections of fixed $c_m$. However, they are defined by  coordinate $x$, and hence we have to relate
the (``bulk'') potential $\psi_{m}$, concentration $c_{m}$ and their derivatives to $x$, which requires further investigation. These functions of $x$ can only be found from the solution of the global problem for the whole channel by imposing  boundary conditions \eqref{c0} and \eqref{psi0}, as well as the zero total current  condition.
Once they are determined, $\phi_{s}(x)$ can be obtained by using
Eq.~\eqref{eq:pot-charge_hs}, and hence  $u_{s}(x)$  from Eq.~\eqref{ui}.

\section{Global diffusio-osmotic flow}
\label{sec:fluxes}

\subsection{General considerations}\label{sec:general}

We focus first to the  expressed per unit channel thickness ion fluxes $\mathcal{J}^{\pm }$ and flow rate $\mathcal{Q}$ of the fluid, which can be written as
\begin{equation}
\mathcal{J}^{\pm }=\frac{1}{2}\int_{-1}^{1}j_{x}^{\pm }dz,\quad \mathcal{Q}=%
\frac{1}{2}\int_{-1}^{1}u_{x}dz,  \label{jq}
\end{equation}%

Integrating Eqs. \eqref{npz} and \eqref{cont_z} and making use of \eqref{jq} we find
\begin{equation}
\frac{H}{2L}\partial _{x}\mathcal{J}^{\pm }+\int_{-1}^{1}\partial
_{z}j_{z}^{\pm }dz=0,  \label{jco}
\end{equation}%
\begin{equation}
\frac{H}{2L}\partial _{x}\mathcal{Q}+\int_{-1}^{1}\partial _{z}u_{z}dz=0.
\label{qco}
\end{equation}%
From impermeability condition (\ref{imper}) it follows that the integrals in Eqs.~\eqref{jco} and \eqref{qco} are zero:
\begin{equation*}
  \int_{-1}^{1}\partial _{z}j_{z}^{\pm
}dz=j_{z}^{\pm }\left( 1\right) -j_{z}^{\pm }\left( -1\right)
= 0,
\end{equation*}
\begin{equation*}
  \int_{-1}^{1}\partial _{z}u_{z}dz=u_{z}\left( 1\right) -u_{z}\left(
-1\right) =0.
\end{equation*}
Thus,
\begin{equation}
\partial _{x}\mathcal{J}^{\pm }=0,  \label{jx}
\end{equation}%
\begin{equation}
\partial _{x}\mathcal{Q}=0.
\label{qx}
\end{equation}%
The last equations imply that $\mathcal{J}^{\pm }$ and $\mathcal{Q}$ given by \eqref{jq} do not vary with $x$, i.e. are the same at any cross-section. This integral representation of the Nernst-Planck and continuity equations can be used to determine the distributions of concentration $c_{m}\left( x\right)$ and potential $\psi _{m}\left( x\right) $ along the channel.

The value of $\mathcal{J}^{\pm }$ may be determined by substituting $j_{x}^{\pm }$ into \eqref{jq} and performing the integration.
These $x-$components of ion fluxes are given by \eqref{jd} and can be rewritten in the dimensionless form as,
\begin{equation}
j_{x}^{\pm }=c^{\pm } \mathrm{Pe} u_{x}+\left( 1\pm \beta \right) \left(
-\partial _{x}c^{\pm }\mp c^{\pm }\partial _{x}\psi \right) ,  \label{NP1}
\end{equation}%
where
\begin{equation}
\mathrm{Pe}=\dfrac{k_{B} T}{2\pi \eta \ell_B \left(
D^{+}+D^{-}\right) }  \label{Pe}
\end{equation}%
is the P\'{e}clet number that characterizes the ratio of the rate of
convection by that of diffusion and the (ion specific) factor $\beta $ is
defined in terms of the difference in diffusion constants of cations and
anions
\begin{equation}
\beta =\frac{D^{+}-D^{-}}{D^{+}+D^{-}},  \label{beta}
\end{equation}%
In such a definition $\beta $ is positive, if cations diffuse faster than
anions, and vice versa. The values of $\beta $ are typically confined
between $-1$ to $1$. For instance, $\beta \simeq 0.285$ for KCH$_{3}$COO, $%
0.014$ for KNO$_{3}$, $-0.207$ for NaCl, and nearly vanishes for KCl~\cite%
{velegol2016}. We also emphasize that for a specific salt $\mathrm{Pe}$ is controlled by  $D^{+}+D^{-}$ and is a constant. Substituting the experimental values for diffusion coefficients~\cite{velegol2016} into Eq.~\eqref{Pe} we obtain for the above inorganic salts $\mathrm{Pe} \simeq 0.31, 0.25, 0.28,$ and 0.23, correspondingly. These values are quite close, and we also note that although $\mathrm{Pe} < 1$, it is not too small to be neglected.

We remark and stress that Eqs.~\eqref{jx} and \eqref{NP1} imply that the zero current condition must refer to the average current $\mathcal{J}^{+}-\mathcal{J}^{-}=0$, as suggested by~\citet{jing2018}, but not to the local values $j_{x}^{+}-j_{x}^{-}$, as proposed by~\citet{keh2005}.

\subsection{``Bulk'' concentration, potential and pressure }\label{sec:equations}

We now turn to the derivation of the equations for $c_m$, $\psi_m$ and $p$ as a function of $x$, which are required to calculate the fluid velocity.

For a thick channel the main contributions to $\mathcal{J}^{\pm }$ and $\mathcal{Q}$ are coming from its
central (``bulk'') part, where the fluid velocity is the sum of the plug diffusio-osmotic and parabolic pressure-driven
velocities:
\[
u_{x} \simeq u_{s}-\frac{1-z^{2}}{2}\partial
_{x}p.
\]%
Consequently, the flow rate that obeys Eq.~\eqref{qx} is given by
\begin{equation}
\mathcal{Q}\simeq u_s-\frac{\partial _{x}p}{3}.
\label{Q}
\end{equation}%
As we have clarified in Sec.~\ref{sec:velocity}, in the general case the slip velocity varies along the channel. Consequently, the constant flow rate can only be provided if the local pressure gradient is non-zero. Moreover, one can argue that since the hydrostatic pressure in both reservoirs are equal, the function $p(x)$ should take its extremum value at some $x$.

Integrating Eq. (\ref{NP1}) over $z$ and using that
\[
\int_{-1}^{1}c^{\pm }u_{x}dz\simeq c_{m}\mathcal{Q},
\]
one obtains
\begin{equation}
\mathcal{J}^{+}\simeq c_{m}\mathrm{Pe}\mathcal{Q}+\left( 1+\beta \right)
\left( -\partial _{x}c_{m}-c_{m}\partial _{x}\psi _{m}\right) ,  \label{jp}
\end{equation}%
\begin{equation}
\mathcal{J}^{-}\simeq c_{m}\mathrm{Pe}\mathcal{Q}+\left( 1-\beta \right)
\left( -\partial _{x}c_{m}+c_{m}\partial _{x}\psi _{m}\right) .  \label{jm}
\end{equation}%

Introducing $\mathcal{J} = \mathcal{J}^{\pm}$, one can exclude $\mathcal{Q}$ by
subtracting these equations to get
\begin{equation}
\partial _{x}\psi _{m}=-\beta \frac{\partial _{x}c_{m}}{c_{m}}.  \label{pot2}
\end{equation}

Substituting Eq.~\eqref{pot2} into Eq.~\eqref{ui} we find that this expression for a diffusio-osmotic slip velocity can be reformulated as
\begin{equation}
u_{s}=-\frac{\partial _{x}c_{m}}{c_{m}} \left[\beta \phi _{s}+4\ln \left[ \cosh \left( \frac{\phi _{s}}{4}\right) \right]\right]. \label{ui2}
\end{equation}
While the form of Eqs.~\eqref{pot2} and \eqref{ui2} is identical to the known formulas for a single wall~\cite%
{prieve1984motion,anderson1989colloid}, the boundary conditions and the derivation itself are different.

Further insight can be gained by integrating Eq.~(\ref{pot2}) and imposing condition (\ref{c0}), which yields
\begin{equation}
\psi _{m}=-\beta \ln c_{m}.  \label{eq:phi0_out}
\end{equation}%
Using this equation we can immediately obtain the midplane potential at the ``salty'' end, which is equal to the voltage $\Delta \psi_m = \psi _{m}(
1) - \psi _{m}(0) $ between reservoirs. The later is given by $\Delta \psi _{m} = -\beta \ln c_{1}$. To get some idea of the orders of magnitude,
with the parameters of Fig.~\ref{fig:pot-charge} we obtain $\Delta \psi _{m} \simeq 1.4$ for NaCl and $-0.1$ for KNO$_{3}$. Clearly, the quantity $
\Delta \psi_m$ should  be  practically zero for KCl since $\beta \simeq 0$.
The form of the above expression for $\Delta \psi _{m}$ is analogous to the known  electrochemistry formula for the diffusion potential between two salt solutions (separated by an infinitesimally thin uncharged membrane) derived for the situation of zero $\mathcal{Q}$. Here we have shown that the same expression  applies when $\mathcal{Q} \neq 0$ too. Returning to   generic Eq.~\eqref{eq:phi0_out}, we emphasize that it describes $\psi_m$ at any cross-section inside the channel, but not just its ends (and is valid at any flow rate of fluid). As a side note, Eq.~\eqref{eq:phi0_out} is similar to the expression describing the potential distribution around
catalytic particles that release ions~\cite{asmolov2022self,asmolov2022COCIS}. However, we are unaware of any prior work that has derived it for a diffusio-osmotic flow in a thick slit.

Substituting (\ref{eq:phi0_out}) into
(\ref{jp}) we obtain an ordinary differential equation
\begin{equation}
\mathcal{J}=c_{m}\mathrm{Pe}\mathcal{Q}-\left( 1-\beta ^{2}\right) \partial
_{x}c_{m}  \label{dc}
\end{equation}%
that provides a route to the determination of
$c_{m}\left( x\right)$.

In the special case of $\mathcal{Q} = 0$ the solution to Eq.~\eqref{dc} satisfying boundary conditions \eqref{c0} yields
\begin{equation}\label{eq:zeroQ}
 \mathcal{J}=-\left( 1-\beta ^{2}\right) \left( c_{1}-1\right)
\end{equation}
and
\begin{equation}
c_{m}=1-\frac{\mathcal{J}}{1-\beta ^{2}}x = 1+\left( c_{1}-1\right) x.
\label{eq:c_lin2}
\end{equation}%
We emphasize that when $\mathcal{Q}$ vanishes,  $ \mathcal{J}$ is negative, i.e. toward the ``fresh'' bath, and the (bulk) concentration $c_m$ augments linearly with $x$.

The emergence of diffusio-osmotic flow, however, normally implies that $\mathcal{Q} \neq 0$. By integrating Eq.~\eqref{dc} and applying the first boundary condition in \eqref{c0} we derive
\begin{equation}
c_{m}=\left( 1-\frac{\mathcal{J}}{\mathrm{Pe}\mathcal{Q}}\right) \exp \left(
\frac{\mathrm{Pe}\mathcal{Q}}{1-\beta ^{2}}x\right) +\frac{\mathcal{J}}{%
\mathrm{Pe}\mathcal{Q}}.  \label{cx}
\end{equation}%
The flux $\mathcal{J}$ can then be found from (\ref{cx}) by using the second boundary condition in \eqref{c0}
\begin{equation}
\mathcal{J}=\mathrm{Pe}\mathcal{Q}\frac{c_{1}-c^{\ast }}{1-c^{\ast }},
\label{JQ}
\end{equation}
where the quantity
\begin{equation}
c^{\ast }=\exp \left( \frac{\mathrm{Pe}\mathcal{Q}}{1-\beta ^{2}}\right)
\label{cc1}
\end{equation}
defines the sign of $\mathcal{J}$, i.e. the direction of ionic flux. Thus $c^{\ast }<1$ for $\mathcal{Q}<0$ that yields a negative $\mathcal{J}$.
However, $c^{\ast }>1$ for $\mathcal{Q}>0$, so the sign of $\mathcal{J}$ is the same as that of $c_{1}-c^{\ast }$. Note that Eq.~\eqref{eq:zeroQ} can be reproduced by expanding \eqref{JQ} about $\mathcal{Q}=0$ and taking the leading order term only. Recall that zero $\mathcal{Q}$ yields  $\mathcal{J}<0$.

We also remark that from Eq.~\eqref{cx} it follows that, if $x \ll 1$, even for a finite flow rate the concentration becomes linear in $x$
\begin{equation}
c_{m}\simeq 1 - \frac{(c_1 - 1) \mathrm{Pe}\mathcal{Q}}{(1 - c^{\ast })(1-\beta ^{2})} x. \label{cx_linear}
\end{equation}%
The same is true close to the ``salty'' end, where $x$ is close to unity
\begin{equation}
c_{m}\simeq c_1 - \frac{(c_1 - 1)c^{\ast } \mathrm{Pe}\mathcal{Q}}{(1 - c^{\ast })(1-\beta ^{2})} (x-1), \label{cx_linear2}
\end{equation}%
but not for the whole channel.

Equation~\eqref{cx} can be regarded as a direct analog of the expression for $c_m$ obtained by \citet{ault.jt:2019} in the assumption that $\phi_s$ and $\mathcal{Q}$ are prescribed.   In our case the surface charge density is fixed, but the surface potential varies along the channel and the flow rate is established self-consistently.
Once $\mathcal{Q}$ is found, $\mathcal{J}$ can be calculated from \eqref{dc} or \eqref{JQ}, and hence $c_m(x)$, $\psi_m(x)$, and $p(x)$ can be obtained.

Since $\mathcal{Q}$ is the same at any cross-section, it
may be determined by integrating Eq.~\eqref{Q} over $x$
\begin{equation}\label{eq:int_Q}
 \mathcal{Q} = \int_{0}^{1} \left[u_s - \frac{\partial _{x}p}{3} \right] dx.
\end{equation}
The second integral in \eqref{eq:int_Q} is  zero since $p(0) =p(1)$. To calculate the first it is convenient to re-expressed $u_s$ given by Eq.~\eqref{ui2} as
\begin{equation}
u_{s}=\mathcal{F}\partial _{x}c_{m},  \label{uis}
\end{equation}
where
\begin{equation}
\mathcal{F}=-\frac{ \beta \phi _{s}+4\ln \left[ \cosh \left( \phi _{s}/4\right) \right] }{c_{m}},  \label{uif}
\end{equation}
is the function of $c_{m}$ solely [see Eq.~\eqref{eq:pot-charge_hs} for $\phi_s$]. Equation~\eqref{eq:int_Q} can  be then transformed to
\begin{equation}
\mathcal{Q}=\int_{0}^{1}\mathcal{F}\partial _{x}c_{m}dx=\int_{1}^{c_{1}}%
\mathcal{F}dc_{m}.  \label{qint}
\end{equation}%
Thus, we have expressed the fluid flow rate in a form of the integral of $\mathcal{F}(c_m)$ over $c_m$, which allows one to treat the diffusio-osmotic problems without tedious and time consuming computations. In essence, all numerical work is now reduced to the trivial calculation of this integral.
In all specimen examples below to obtain $\mathcal{Q}$ we will integrate Eq.~\eqref{qint} numerically (or analytically) using $\phi_s$ calculated from Eq.~\eqref{eq:pot-charge_hs}.

From \eqref{qint}
it follows that $\mathcal{Q}$ is controlled only by $\ell_{GC}$ [that impacts $\phi _{s}$] and $c_1$, but does not depend on the
P\'{e}clet number or distributions of  concentration (\ref{cx})
and pressure along the channel. In other words, to attain the high flow rate we need highly charged surfaces and a large  concentration drop
between reservoirs.
The case of $\mathcal{Q}=0$ is also of much interest.
It is evident that the situation of zero $\mathcal{Q}$ can arise from two rather different mechanisms. The first, which occurs for the uncharged surfaces ($\phi_s = 0$),  provides $\mathcal{F} \equiv 0$, so that the diffusio-osmotic flow is not induced at all. The flux of ions throughout the channel involves only their  diffusion and migration. The second, which occurs for charged surfaces, involves the sign reversal of   $\mathcal{F}$. We discuss this case further. The integral in \eqref{qint} could becomes zero, if the function $\mathcal{F}$ is  alternating, which implies that $\mathcal{F}=0$ has a root in the interval of $c_m$ from 1 to $c_1$. It follows from Eq.~\eqref{uif} that it exists only for a negative $\beta \phi_s$. It is possible to determine the value of $c_1$ for a given $\ell_{GC}$ (or vice versa) for which zero flow rate occurs by means of a construction analogous to the Maxwell one (of equal areas) employed for phase coexistence.

\begin{figure}[h]
\begin{center}
\includegraphics[width=1\columnwidth]{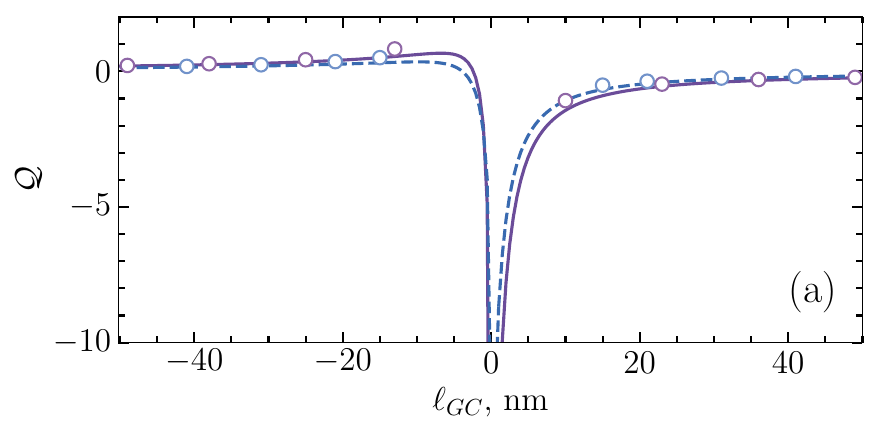}\\
\includegraphics[width=1\columnwidth]{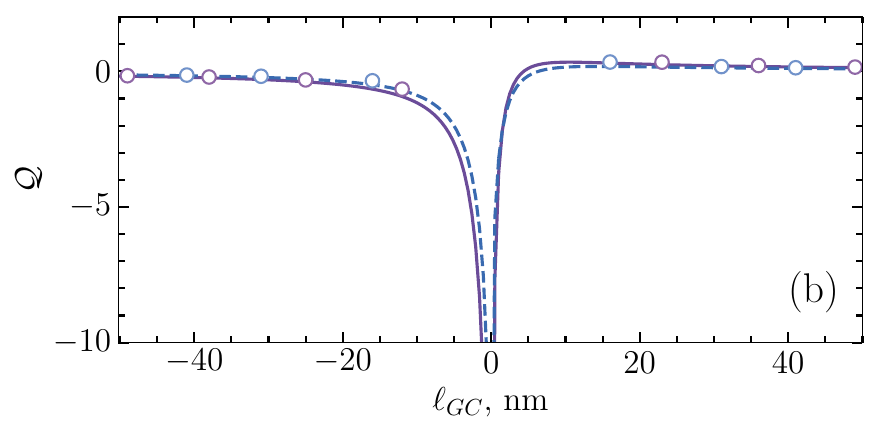}
\end{center}
\caption{$\mathcal{Q}$ as a function of $\ell_{GC}$ computed using $c_1 =
10^3$ and 10 (solid and dashed curves) for KCH$_{3}$COO [$\beta = 0.286$] (a) and NaCl [$\beta = -0.208$] (b).
Circles show calculations from Eq.~\eqref{eq:q_lin}. }
\label{fig:qb}
\end{figure}

By varying $\ell_{GC}$ and calculating $\phi_s$ it is possible to obtain the curves for $\mathcal{Q}$ shown in Fig.~\ref{fig:qb}(a). The calculations
are made using $c_1 = 10^3 $ and 10, and for this example we fix $\beta = 0.286$ of KCH$_{3}$COO.
For positively charged surfaces the flow rate is negative [since the numerator in \eqref{uif} is positive] and increases monotonically  with $\ell_{
GC}$ by slowly approaching zero at large values (i.e. at small surface charge densities). Since for negatively charged surfaces the numerator in \eqref{uif} can take any sign depending on
the value of $\phi_s$, the flow rate could be either positive or negative (or zero). On increasing $\ell_{GC}$ (reducing $|\ell_{GC}|$) $\mathcal{Q
}$ increases nonlinearly (from zero), exhibits a maximum and then decreases sharply to a negative value becoming rather large in magnitude. It is evident that for a given $\mathcal{Q}$ there are always two possible  $\ell_{GC}$ (and hence $\phi_s$), so a great care should be taken when the measured fluid flow rate is used to infer the surface charge/potential~\cite{ault.jt:2019,lee.c:2014}.
Now recall that $\mathcal{F}$ could turn to zero,  or, equivalently, $\beta \phi _{s} + 4\ln
\left[ \cosh \left( \phi _{s}/4\right)\right] = 0$, only for negative $\beta \phi_s$. With $\beta = 0.286$ the root is $\phi_s \simeq -2.4$. At this (sufficiently large) surface potential the electro- and chemiosmotic flow rates are significant, but they cancel out each other.
Finally, we remark that the curves corresponding to different values of $c_1$ are of the same shape, but $\mathcal{Q}$ is slightly smaller in magnitude for smaller $c_1$. Thus,  the data in Fig.~\ref{fig:qb}(a) supports the above argument about an influence of $c_1$ on $\mathcal{Q}$. Meantime, it can be seen that a key player in tuning the flow rate is the surface charge density, but not the concentration drop.
In essence, the effect of concentration drop is quite small. The nature of this apparently unexpected result is more or less clear. The point is that the main contribution to the integral \eqref{qint} comes from the region of lower $c_m$ (and simultaneously higher $\phi_s$), where the slip velocity should be large. We return to this point in Sec.~\ref{sec:slip}.
Finally, we note that at $\ell_{GC} \to 0$ the surface potential diverges and this is accompanied by a divergence of $\mathcal{Q}$.  In practice, however, it is unlikely that $|\ell_{GC}|$ could be  smaller than 1 nm~\cite{vinogradova.oi:2024}. Below we will use this value as a lower bound to the  Gouy-Chapman length.

The above results for $\mathcal{Q}$ refer to fixed positive $\beta$ [see Fig.~\ref{fig:qb}(a)]. If we keep the same values of parameters, but fix (negative) $\beta = -0.208$ of NaCl, we move to the situation displayed in Fig.~\ref{fig:qb}(b). Here the maximum is less pronounced and occurs at positive $\ell_{GC}$. The negative values  of $\mathcal{Q}$ can be attained  for both negative and positive $\ell_{GC}$, whereas the flow towards ``salty'' reservoir can now emerge only if surfaces are  positively charged.

\begin{figure}[h]
\begin{center}
\includegraphics[width=1\columnwidth]{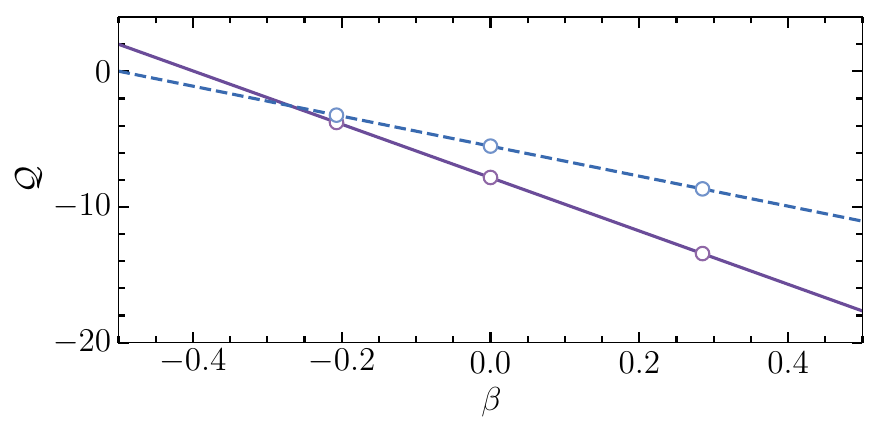}
\end{center}
\caption{The flow rate $\mathcal{Q}$ as a function of $\beta$ computed using $\ell_{GC} = 1$ nm for $c_1 = 10^3$ and 10 (solid and dashed curves). Circles from left to right mark values of $\mathcal{Q}$ that correspond to NaCl, KCl, and KCH$_{3}$COO. }
\label{fig:q_beta}
\end{figure}

Figure~\ref{fig:q_beta} shows $\mathcal{Q}$  obtained at fixed $\ell_{GC}$, but varying $\beta$. The calculations are
 made using $\ell_{GC} = 1$ nm and the same two values of $c_1$ as in Fig.~\ref{fig:qb}. It can be seen that the flow rate decreases linearly with $\beta$ and the (negative) slope $(\partial \mathcal{Q}/\partial \beta)_{\ell_{GC}}$ is larger for higher $c_1$. Figure~\ref{fig:q_beta} is also intended to illustrate that depending on $\beta$, the flow rate can be either positive or negative, or zero. With given parameters  the zero of $\mathcal{Q}$ that corresponds to $c_1 = 10$ and $10^3$ is attained when $\beta = -0.498$ and -0.398. We have marked with symbols the flow rates that correspond to some standard salts. Recall that $\beta$ for KCl is nearly zero, which means  that electroosmosis does not emerge. However, $\mathcal{Q}$ is large and negative, thanks to chemiosmotic flow, from  ``salty'' towards ``fresh'' reservoir. For NaCl the flow rate is smaller in magnitude, but for KCH$_{3}$COO larger. The reason for this difference is clear. In the first case the emerging electroosmotic flow results in a decrement in a total $\mathcal{Q}$ compared to the case of zero $\beta$, but in the second - in an increment. For these three salts the flow rate is negative, but this does not mean that a positive $\mathcal{Q}$ cannot be attained. For example, the flow towards ``salty'' bath is expected for  NaOH (since its $\beta \simeq -0.596$~\cite{velegol2016}).

\begin{figure}[h]
\begin{center}
\includegraphics[width=1\columnwidth]{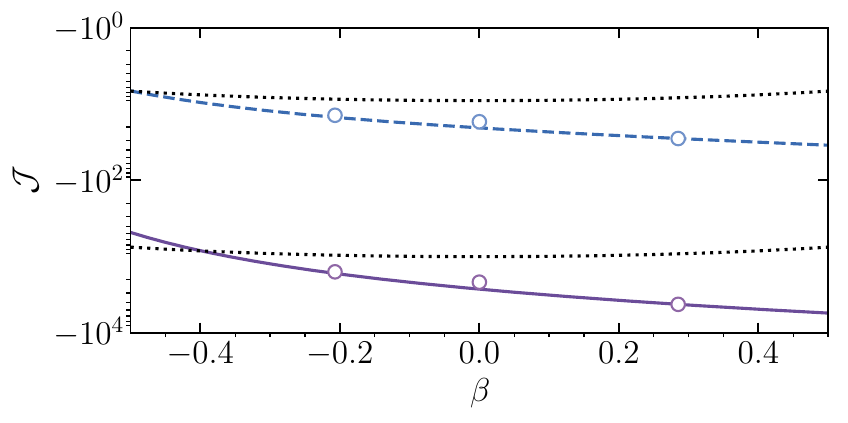}
\end{center}
\caption{$\mathcal{J}$ vs. $\beta$ calculated with the same parameters as in Fig.~\ref{fig:q_beta}
using $\mathrm{Pe} = 0.31$ (solid and dashed curves). Circles from left to right show $\mathcal{J}$ for NaCl, KCl, and KCH$_{3}$COO. Dotted curves are obtained using Eq.~\eqref{eq:zeroQ}. }
\label{fig:j_beta}
\end{figure}

Figure \ref{fig:j_beta} illustrates the ion flux $\mathcal{J}$ that corresponds to the flow rates of fluid presented in  Fig.~\ref{fig:q_beta}. The curves are calculated from Eq.~\eqref{JQ} using $\mathrm{Pe} = 0.31$ obtained for KCH$_{3}$COO. However, only the rightmost circles refer to KCH$_{3}$COO; the curves are provided simply as a guide for the eye. The left circles and the center ones refer to NaCl and KCl, and calculated using their $\mathrm{Pe} = 0.28$ and 0.23, correspondingly. It can be seen that they are well fitted by the theoretical curves obtained at slightly higher  $\mathrm{Pe}$. With these parameters $\mathcal{J}$ is negative and  monotonically decreases [but increases in magnitude]  with $\beta$. It is also well seen that for a larger value of $c_1$ the magnitude of $\mathcal{J}$ is higher.
Also included are calculations from Eq.~\eqref{eq:zeroQ}, which predicts that $\mathcal{J}$  takes its minimum value of $1-c_1$ at $\beta = 0$. It is evident that there is always an intersection of the both curves for a given $c_1$. In this case $\mathcal{Q}=0$, so that there is no convective transfer of ions. The examples of salts in Fig.~\ref{fig:j_beta} correspond to a negative $\mathcal{Q}$, and it is well seen that in this situation the fluid flow enhances significantly the ionic flux in the direction of ``fresh'' bath.

\begin{figure}[h]
\begin{center}
\includegraphics[width=1\columnwidth]{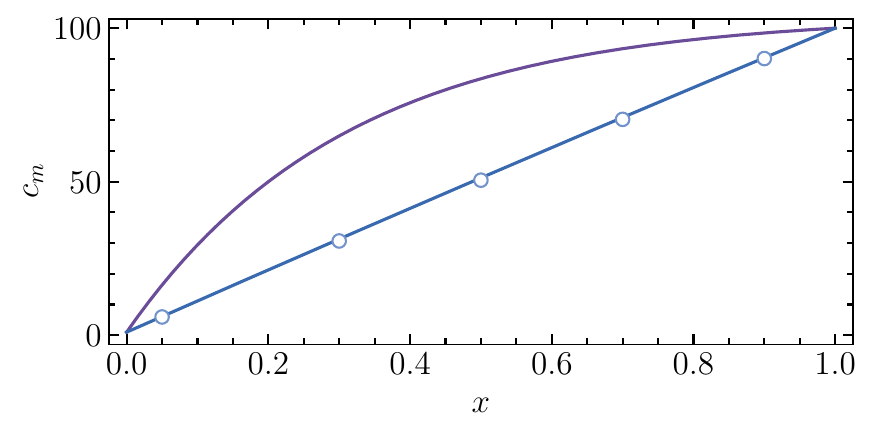}\\[0pt]
\end{center}
\caption{Concentration $c_m$ as a function of $x$
for NaCl [$\beta = -0.207$, $\mathrm{Pe} = 0.28$] computed using $c_1 = 10^2$ and $\ell_{GC} = -1$ and 5 nm (solid curves from top to bottom). Circles show calculations from Eq.~\eqref{eq:c_lin2}. }
\label{fig:c}
\end{figure}

The calculations of $\mathcal{Q}$ and $\mathcal{J}$ opened us a direct route to the determination of the concentration $c_m$ distribution  along
the channel. The local $c_m$ can be determined from Eq.~\eqref{cx}, which reduces to Eq.~\eqref{eq:c_lin2} when $\mathcal{Q} = 0$.  Figure~\ref{fig:c} includes theoretical curves calculated for NaCl. The calculations are made using several $\ell_{GC}$ from $\pm 1$ to $\pm 20$ nm and $c_1 =
10^2$, but the results are shown only for $\ell_{GC} = -1$ and $5$ nm  since they constrain all other curves obtained. The concentration
distributions are generally convex, that is the deviations from Eq.~\eqref{eq:c_lin2}, if any, are always in the direction of larger $c_m$, and
note that they increase on reducing $|\ell_{GC}|$. 

For $\ell_{GC} \geq 3$ nm the concentration curves practically merge into a straight line given
by  \eqref{eq:c_lin2}. An explanation can be obtained if we invoke Eq.~\eqref{cx}. By differentiating it twice in respect to $x$ we find that it is
proportional to $\mathcal{Q}^2 [1 - \mathcal{J}/(\mathrm{Pe} \mathcal{Q})]$. Thus with positive $\mathcal{J}/\mathcal{Q}$ the function $c_m (x)$ is
convex, if $\mathcal{Q} \neq 0$, and zero curvature is expected when the flow rate approaches to zero. The curves in Fig.~\ref{fig:c} correspond to
different values of $\mathcal{Q}$ and $\mathcal{J}$. On the upper curve, $\mathcal{Q}\simeq -11.0$ and $\mathcal{J}\simeq -321.2$, but for the
lower one $\mathcal{Q}\simeq 0.0$ and $\mathcal{J}\simeq -94.5$. So, in the former case, the flow rate is large and negative, so is $\partial_{xx} c
_m$. In the latter case $\partial_{xx} c_m \simeq 0$ in virtue of nearly zero $\mathcal{Q}$. Note that since the Gouy-Chapman length has immense
variability depending on wall material and surface modification, and since $\beta$ for some specific electrolyte can, in principle, take any value
from -1 to 1, the global picture becomes rather rich.
Here we make no attempt to investigate all possible scenarios for  profiles of $c_m$, which deserves a separate publication. Instead, we simply intended  to illustrate how to interpret the shape of the concentration curves using NaCl as an example.

 \begin{figure}[h]
\begin{center}
\includegraphics[width=1\columnwidth]{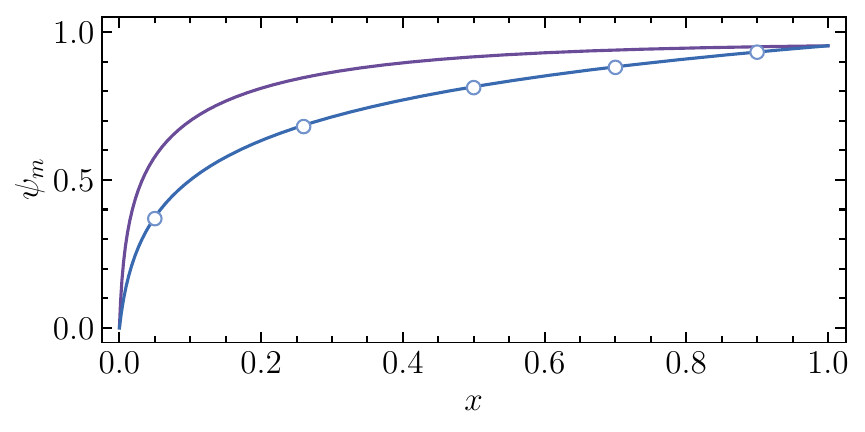}
\end{center}
\caption{Potential $\psi_m$ as a function of $x$ computed from Eq.~\eqref{eq:phi0_out} for the concentration profiles of NaCl displayed in Fig.~\ref{fig:c}. }
\label{fig:Psim}
\end{figure}

The profiles of $\psi_m$ along the channel are of interest. The calculations of local midplane potentials for the concentration distributions shown
in Fig.~\ref{fig:c} are made using Eq.~\eqref{eq:phi0_out} and included in Fig.~\ref{fig:Psim}. The results of calculations indicate that while
convection does not affect $\Delta \psi_m$ for NaCl, it augments the local values of $\psi_m$ inside the channel, as well as $\partial_x \psi_m$ at
sufficiently small $x$. It can also be seen that the function $\psi_m$ increases more rapidly with $x$ close to the ``fresh'' end of the channel
than near the ``salty'' reservoir. This implies that the slip velocity monotonically reduces in
magnitude with $x$ [as follows from Eqs.~\eqref{pot2} and \eqref{ui2}]. Eq.~\eqref{Q} clarifies that in such a situation $\partial_x p$ takes non-
zero values within the channel in order to provide the same flow rate of fluid in every cross-section.

\begin{figure}[h]
\begin{center}
\includegraphics[width=1\columnwidth]{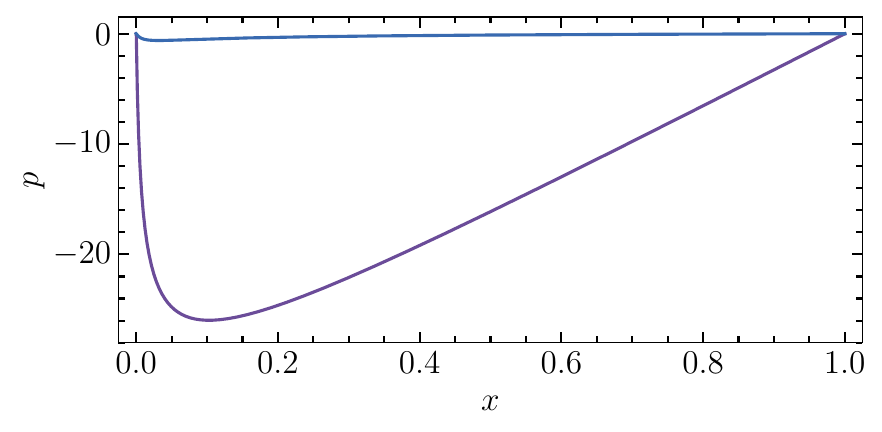}
\end{center}
\caption{The distribution of local hydrostatic pressure $p$ along the channel computed from Eq.~\eqref{px} for the concentration profiles of NaCl shown in Fig.~\ref{fig:c}. From top to bottom $\ell_{GC} = 5$ and $-1$ nm.}
\label{fig:p}
\end{figure}

The local pressure can be obtained by integrating Eq.~\eqref{Q} from $x=0$ to an arbitrary $x \leq 1$:
\begin{equation}
p\left( x\right) =3\left( \int_{1}^{c_{m}\left( x\right) }\mathcal{F}\left(
c_{m}\right) dc_{m}-\mathcal{Q}x\right) .  \label{px}
\end{equation}
The results of calculations are illustrated in Fig.~\ref{fig:p}. For this specimen example we again use the concentration profiles for NaCl displayed in Fig.~\ref{fig:c}. We begin with the lower curve computed using $\ell_{GC} = -1$ nm. As $x$ increases from 0, the pressure first decreases very rapidly and after taking its minimum value begins to augment (more slowly and linearly above $x \simeq 0.2$). This implies that the pressure gradient $\partial_x p$ is non-uniform, i.e. depends on $x$. Closer to the ``fresh'' reservoir it is large and negative, but at some $x$ reverses its sign and finally becomes both positive and constant. There have been prior reports on such a form of the pressure curve~\cite{ault.jt:2019,peters2016}.
It is also seen that for a channel of $\ell_{GC} = 5$ nm (the upper curve in Fig.~\ref{fig:p}), where $\mathcal{Q} \simeq 0$, the minimum of $p$ is much less pronounced (and shifted to smaller $x$). The local pressure itself is extremely small and, in essence, one can consider that it does not arise at all, $p(x) \simeq 0$.

\subsection{Velocity profiles and apparent slip}\label{sec:slip}

\begin{figure}[h]
\begin{center}
\includegraphics[width=1\columnwidth]{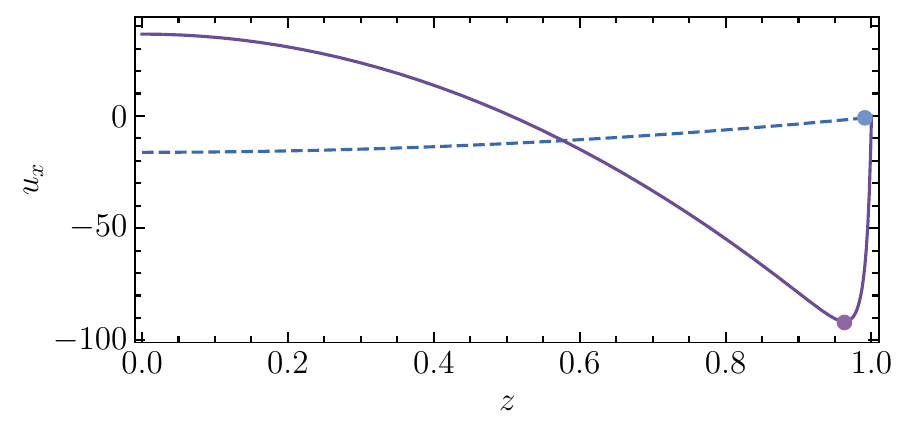}
\end{center}
\caption{Velocity profiles $u_x(z)$  computed for the lower curve in Fig.~\ref{fig:p} that refer to cross-sections $x=0.02$ and $0.5$ (the solid and
dashed curves). Circles show calculations from Eq.~\eqref{uis}. }
\label{fig:u}
\end{figure}

The non-uniform pressure gradient $\partial_x p$ gives rise to a supplementary flow  that has significant repercussions for the fluid velocity
profile $u_x(z)$. The fluid velocity  becomes a linear superposition of two terms: a  plug diffusio-osmotic flow
and the familiar parabolic profile of Poiseuille flow  between parallel
(no-slip) planes. Figure~\ref{fig:u} includes theoretical curves calculated for  $x=0.02$ and $0.5$ with the parameters of the lower curve in Fig.~
\ref{fig:p}. For these cross-sections $c_m \simeq 6.1$ and $77.0$, correspondingly, but pressures are quite close ($p \simeq -20.6$ and $-16.1$).
The velocity profiles are calculated from Eq.~\eqref{vel} with the local mobilities given by \eqref{uh}, \eqref{ueo}, and \eqref{udo2}. The
local gradients are found from Eqs.~\eqref{pot2}, \eqref{cx}, and \eqref{px}.
At $x=0.02$ the fluid velocity plotted against $z$ has a complex rippled shape with a ``W-profile'' (which is dubbed here the
wimple  by analogy to
a term used to describe a rippled deformation of liquid drops~\cite{clasohm.ly:2005}). This includes a large central region, where the velocity
profile is convex, and an area  in the neighborhood of the walls with the concave profile. Although the global flow rate $\mathcal{Q}$ is toward
the ``fresh'' reservoir, the very central part of fluid (near the midplane) moves away from it.
The velocity reverses its sign from positive close to the midplane to negative, and after taking a minimum increases to zero at the wall. The
negative area prevails since for this specific channel $\mathcal{Q} = \int_{0}^{1} u_x(z) dz \simeq -11$ as discussed above. At larger $x$ this shape evolves into a conventional (concave) parabola (or ``U-profile''). Our theory  allows us to make contact with the simulations, which  reported similar qualitative features of the evolution of velocity profile in the channel~\cite{ryzhkov.ii:2018} and provides a direct physical explanation of this result.  Since the pressure gradient reverses its sign inside the channel (see Fig.~\ref{fig:p} ), the arising Poiseuille flow can be toward either ``fresh'' or ``salty'' reservoir depending on $x$. As $x=0.02$ the pressure-driven flow is directed  oppositely to the diffusioosmosis.  As a result,  in the vicinity of the ``fresh'' end we observe the wimple formation. The wimple amplitude decreases with $x$, and when the pressure gradient vanishes the velocity profile becomes flat. At larger $x$ both flows are co-directed leading to the conventional parabolic profile. Its amplitude augments up to $x \simeq 0.2$. On increasing $x$ further the velocity profile remains the same for every cross-section. As a side note, the analogous picture is observed if we make calculation with the parameters of the upper curve in Fig.~\ref{fig:p}, where $\mathcal{\mathcal{Q}} \simeq 0$, but $u_x(z)$ becomes much smaller in magnitude.

\begin{figure}[h]
\begin{center}
\includegraphics[width=1\columnwidth]{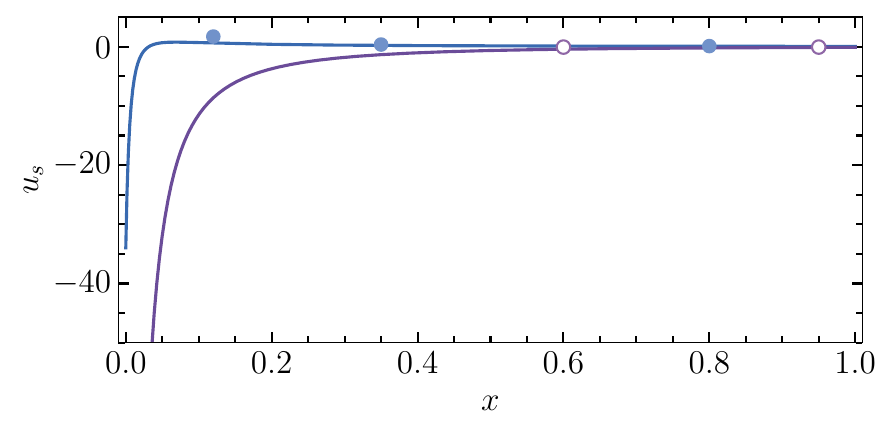}
\end{center}
\caption{Slip velocity $u_s$ vs $x$ computed from Eq.~\eqref{uis} for the concentration profiles of NaCl displayed in Fig.~\ref{fig:c}. From top to bottom $\ell_{GC} = 5$ and $-1$ nm. Filled circles show calculations from Eq.~\eqref{u_DB1}. Open circles show predictions of Eq.~\eqref{u_DB} with $c_m$ calculated from \eqref{cx_linear2} and $\partial_x c_m$ given by \eqref{eq:der_largex}.}
\label{fig:us}
\end{figure}

We now turn to the apparent diffusio-osmotic slip velocity that can be calculated from Eq.~\eqref{uis}. In Fig.~\ref{fig:u}, we have marked with
circles the quantities $u_s$ for the corresponding curves. The locus of  $u_x(z)=u_s$ is at  some distances of the order of the local Debye length [i.e. $O(\lambda_D^{\star} c_m^{-1/2})$] from the walls. Note that for the wimple profile the velocity $u_x(z)$ takes here its  minimum value. By
varying $x$ and performing the same calculations we obtain the curves for $u_s$ shown in Fig.~\ref{fig:us}. They again refer to the concentration
profiles of NaCl displayed in Fig.~\ref{fig:c}. An overall conclusion from this plot is that the slip velocity $u_s$  significantly varies along
the channel. It is quite large in magnitude in the vicinity of the ``fresh'' reservoir, but much smaller close to the ``salty'' one
[and this, in fact, is the main reason  for a weak dependence of $\mathcal{Q}$ on $c_1$ in Fig.~\ref{fig:qb}]. For the lower curve that refers to $\mathcal{Q} \simeq -11$ the negative slip velocity monotonically reduces in magnitude with $x$. We emphasize that even for a
channel, which is normally not regarded as highly charged and of $\mathcal{Q} \simeq 0$, there emerges a finite diffusio-osmotic slip that has been
usually swept under the carpet. Indeed, the (upper) curve corresponds to zero $\mathcal{Q}$, but it can be seen that $u_s$ is finite and alternating. The slip velocity is large and negative near the ``fresh'' reservoir, but small and positive in most of the channel. The nature of this is more or less apparent now. At small $x$ the surface potential is large, so the negative chemiosmotic term in Eq.~\eqref{uif} dominates. On increasing $x$ the surface potential reduces and the positive electroosmotic term becomes the leading. The integral contributions of these two terms of the opposite sign to the flow rate $\mathcal{Q}$ given by Eq.~\eqref{qint} are finite, but they cancel out.

Finally, we present some analytical calculations for the case of low surface potential. From Eq.~\eqref{ui} it follows that when $\phi_s \leq 1$
the electroosmotic contribution to the diffusio-osmotic velocity (the first term) is $O\left( \phi _{s}\right)$, while the
chemiosmotic one (the second term) is $O\left( \phi _{s}^{2}\right).$ Consequently, the latter can be safely
neglected. Then using Eq.~\eqref{fs_DB} the slip velocity can be approximated by%
\begin{equation}
u_{s}\simeq -2\beta \frac{\lambda _{D}^{\star }\partial _{x}c_{m}}{%
c_{m}^{3/2}\ell _{GC}}.  \label{u_DB}
\end{equation}%

By substituting the last equation into (\ref{qint}) we obtain the flow rate for this specific case%
\begin{equation}
\mathcal{Q}=-4\beta \frac{\lambda _{D}^{\star }}{\ell _{GC}}\left(
1-c_{1}^{-1/2}\right).  \label{eq:q_lin}
\end{equation}%
The relative contribution of the concentration drop to the fluid flow rate is determined by the second term in the brackets. It can be seen that the latter is small, $O(c_1^{-1/2})$, which explains a weak dependence of $\mathcal{Q}$ on $c_1$. From Eq.~\eqref{eq:q_lin} it follows that $\mathcal{Q}$ is negative, if $\beta/\ell _{GC}$ is positive, and vice versa. The calculations from \eqref{eq:q_lin} are included in Fig.~\ref{fig:qb}. The fits are quite good for large $|\ell _{GC}|$ (small $|\phi_s|$), i.e. within the range of validity of this equation.

For $\mathcal{Q} \leq 1$ Eqs.~(\ref{cx})-(\ref{cc1}) can be expanded about $\mathcal{Q} = 0$ and, to first order, one obtains Eq.~\eqref{eq:c_lin2}  for $c_m$. Thus, the linear in $x$ concentration distribution with $\partial _{x}c_{m} \simeq c_1 - 1$ should be a sensible approximation. Substituting it to \eqref{u_DB} we find that the slip velocity is given by
\begin{equation}
u_{s}\simeq -2\beta \frac{\lambda _{D}^{\star }}{\ell _{GC}} \frac{ c_{1}-1 }{\left[
1+\left( c_{1}-1\right) x\right] ^{3/2}}.  \label{u_DB1}
\end{equation}
The calculations from Eq.~\eqref{u_DB1} are compared with the results for $\mathcal{Q} \simeq 0$ presented in Fig.~\ref{fig:us} (the upper curve). It can be seen that \eqref{u_DB1} provides a very good fit to the numerical curve  down to $x \simeq 0.1$. At  smaller $x$ the surface potential becomes large, so this simple analytical equation does not apply.

We remark and stress that even for small surface potentials and constant concentration gradient $u_s$ remains a non-linear function of $x$. The constant slip velocity
\begin{equation}
u_{s}\simeq -2\beta \frac{\lambda _{D}^{\star }}{\ell _{GC}} ( c_{1}-1 ) \ll 1  \label{u_DB1_small}
\end{equation}
can be expected only if the concentration drop is extremely small, $c_1 -1 \ll 1$.

For a finite $\mathcal{Q}$ the concentration gradient is generally non-uniform, but becomes a constant when $1-x$ is sufficiently small. In this situation $\phi_s$ becomes small too and $c_m$ may be determined from Eq.~\eqref{cx_linear2}. Differentiating this equation with respect to $x$ we obtain
\begin{equation}\label{eq:der_largex}
 \partial_x c_m \simeq - \frac{(c_1 - 1)c^{\ast } \mathrm{Pe}\mathcal{Q}}{(1 - c^{\ast })(1-\beta ^{2})}
\end{equation}
The slip velocity $u_s$ can then be obtained by substituting Eqs.~\eqref{cx_linear2} and \eqref{eq:der_largex} into \eqref{u_DB}. Such calculations are included in Fig.~\ref{fig:us} and we see that they are in excellent agreement with the numerical results for a branch of the lower curve that corresponds to sufficiently small $\phi_s$ (and a large flow rate of fluid).

\section{Concluding remarks}\label{sec:concl}

We proposed a theory of a diffusio-osmotic flow in a thick planar channel, which is valid  even when the surface potential and charge density are quite large and applies at any concentration drop. The theory provides considerable insight into physics of diffusio-osmosis and has the merit of being very well suited to numerical work by dramatically simplifying it compared to prior approaches that involve a numerical solution to the system of partial differential equations. Our approach  also yielded some useful analytical relations that have never been reported, and we believe we have provided satisfactory answers to several long-standing questions formulated at the beginning of the paper. Thus, our results  may be useful for testing numerical approaches or for predicting data that still cannot be   obtained from experiment.

The main results of our work can be summarized as
follows. We have derived an analytical formula that relates the local ``bulk''  concentration $c_m(x)$ to the flow rate of fluid $\mathcal{Q}$ and depends on the P\'{e}clet number and diffusivities of ions. In general this varies non-linearly along the channel, that is the concentration gradient $\partial_x c_m$ is not a constant. The slip velocity $u_s$ is also non-uniform, so is the pressure gradient that arises  naturally in our analysis  providing an equal flow rate in every cross-section. We have emphasized that at a finite flow rate $u_s$ decreases monotonically in magnitude from its largest value near the ``fresh'' reservoir down to the smallest in the vicinity of the ``salty'' one. Unexpectedly, our analysis has revealed that zero $\mathcal{Q}$ does not necessarily imply the absence of the diffusio-osmotic flow, and can be attained when  $u_s$ becomes alternating. Finally, our exact analytical solution for the local diffusio-osmotic mobility opened a route to the determination of the local velocity profiles $u_x(z)$. We have clarified the evolution of their form along the channel associated with the related pressure changes.

Certain aspects of our work warrant further comments. The flow rate $\mathcal{Q}$ is an important quantity in our theory because, in essence, it sets everything. Once $\mathcal{Q}$ is determined, the ionic flux $\mathcal{J}$, local ``bulk'' concentrations $c_m$ and potentials $\psi_m$, and hence $u_s$ can be easily obtained by using
the analytical expressions derived here. We have formulated the expression for the flow rate of fluid in the form of a simple integral that can easily be computed. It has been shown that $\mathcal{Q}$ depends neither on the P\'{e}clet number, nor on the distribution of local concentrations along the channel  being controlled   only by the surface charge and (to a less extent) the concentration drop. Our identification of two separate mechanisms for  vanishing of $\mathcal{Q}$ should also be of much significance. The first, associated with zero surface potential (or charge), is, of course, well-known. The second mechanism occurs for charged surfaces and requires the alternating integrand in the equation for $\mathcal{Q}$. Thus, both chemi- and electroosmotic flows emerge, but their integral contributions cancel out.

One of the aims of our study was to provide fundamental theoretical understanding of some important features of diffusion-osmotic flow in a channel, which cannot be clarified from experiment, but are crucial for  interpreting the data or predictive purpose. In addition to  mentioned above these include such issues as the direction of the flow rate of fluid and that of the flux of ions.
In particular we found that for a highly charged channel $u_s$ is toward ``fresh'' reservoir, if $\beta \ell _{GC}$ is positive. For a finite surface charge and negative $\beta \ell _{GC}$ the slip velocity can be toward any reservoir and become even alternating. Our analysis showed that the ion flux is related to the flow rate  of fluid and we clarified when and why $\mathcal{J}$ and $\mathcal{Q}$ become co-directed, i.e. found the conditions for the convective enhancement of $\mathcal{J}$. For example, the ion flux can be significantly enhanced by the fluid flow, if the latter is toward the ``fresh'' reservoir and $\beta \ell _{GC}$ is positive.

  A diffusio-osmotic flow is at the origin of  migration of charged particles induced by salt gradient and termed diffusiophoresis. This phenomenon represents an efficient means to efficiently manipulate the charged colloids, by inducing their spreading and focusing~\cite{abecassis.b:2009,arya.p:2021,ebel.jp:1988,wanunu.m:2010}. Since for particles, which  radius significantly exceeds the Debye length, the same equation for $u_s$ applies~\cite{nourhani2015,asmolov2022COCIS,ault2024},   our results may be of help at improving the description of diffusio-phoretic phenomena and devices.

The solutions derived here may also be useful for some electrochemistry  areas, such as related to permeable membranes.
One problem for which our work is relevant is that of the diffusion (or liquid junction)  potential $\Delta \psi_m$ associated with many practical aspects of
reference electrodes, which cannot be measured directly and is difficult to interpret since the stationary state is non-equilibrium~\cite{tsirlina.ga:2013}. Electrochemists have long treated this potential as occurring at zero fluid flow rate. Our results show  that the diffusion potential is not specific to the magnitude of $\mathcal{Q}$, pointing out that the classical equation for $\Delta \psi_m$ has validity beyond its initial assumptions. However, the real membranes are porous materials of a finite thickness and we found that the local $\psi_m$ inside their pores is significantly affected by convection. This might have a repercussion for various applications.
The same concerns the effects of a local diffusio-osmotic flow coupled with zero total flow rate that have never been considered in membrane science, but their implications could be large.

The salinity gradient has also been proposed as a source of an electric current generation~\cite{siria.a:2013,cipollina.a:2016}, although  there are still some doubts about the efficiency of this~\cite{elimelech.m:2024}. Our theory of diffusio-osmosis cannot be immediately applied to the description and optimization of electric current since we have addressed so-called an  ``open circuit'', where there is no electronic connection through the wires (and hence the current is zero). The approach we reported, however, could be extended for a ``closed circuit'' configuration. Our calculations are currently in progress for this.

As mentioned in the introduction, another avenue for driving flow on these scales is to exploit hydrodynamic slip. The later can yield considerably enhanced electro-osmotic flow in the channel (see for example \cite{vinogradova.oi:2023}, and references therein), as well as affects diffusio-osmosis~\cite{ajdari.a:2006}. It would be of some interest to examine the diffusio-osmotic flow between two planes by imposing the hydrodynamic slip boundary conditions. A systematic study of this kind would constitute a significant extension of our work.
Another fruitful direction would be to consider diffusion-osmosis flow in channels with a non-uniform surface charge density, including anisotropic ones. While the effective electro-osmotic slip in such systems is well investigated~\cite{anderson.jl:1985,ajdari.a:1995,belyaev.av:2011}, the quantitative understanding of the diffusio-osmotic one is still challenging.

\begin{acknowledgments}

This work was supported by the Ministry of Science and Higher Education of the Russian Federation.
\end{acknowledgments}

\section*{DATA AVAILABILITY}

The data that support the findings of this study are available within the
article.

\section*{AUTHOR DECLARATIONS}

The authors have no conflicts to disclose.

%

\end{document}